\def\pmb#1{\setbox0=\hbox{#1}%
   \kern-.025em\copy0\kern-\wd0
   \kern.05em\copy0\kern-\wd0
   \kern-0.025em\raise.0433em\box0}
\def\gta{\mathrel{{\lower 3pt\hbox{$\mathchar"218$}}\hskip-8pt
   \raise 2pt\hbox{$\mathchar"13E$}}}
\def\lta{\mathrel{{\lower 3pt\hbox{$\mathchar"218$}}\hskip-8pt
   \raise 2pt\hbox{$\mathchar"13C$}}}
\def\half{{\scriptstyle{1\over2}}}
\def\dagg{\phantom{\dagger}}            % This lines subscripts up better
\def\bfnabla{\bf \pmb{$\nabla$}\/}

\def\today{\number\day\space\ifcase\month\or
  January\or February\or March\or April\or May\or June\or
  July\or August\or September\or October\or November\or December\fi
 \space\number\year}
\documentstyle[epsfig]{aipproc}

\begin{document}
\title{Stripes, Carriers, and High-\pmb{$T_c$}\/ in the Cuprates} 

\author{Josef Ashkenazi}
\address{Physics Department, University of Miami, P.O. Box 248046, Coral
Gables, FL 33124, U.S.A.} 

%\lefthead{LEFT head}
%\righthead{RIGHT head}
\maketitle

\begin{abstract}
Considering both ``large-$U$'' and ``small-$U$'' orbitals it is found
that the high-$T_c$ cuprates are characterized by a striped structure,
and three types of carriers: polaron-like ``stripons'' carrying charge,
``quasielectrons'' carrying charge and spin, and ``svivons'' carrying
spin and lattice distortion. It is shown that this electronic structure
leads to the anomalous physical properties of the cuprates, and
specifically the systematic behavior of the resistivity, Hall constant,
and thermoelectric power. High-$T_c$ pairing results from transitions
between pair states of quasielectrons and stripons through the exchange
of svivons. A pseudogap phase occurs when pairing takes place above the
temperature where stripons become coherent, and this temperature
determines the Uemura limit. 
\end{abstract}

\section*{Introduction}

The existence of static stripes in the CuO$_2$ planes has been observed
in some superconducting cuprates \cite{Bian,Tran}, and there is growing
evidence on the existence of dynamic stripes in others \cite{stripes}.
Many experimental observations have been pointing to the presence of
both itinerant and almost localized (or polaron-like) carriers in these
materials. 

Though one-band theoretical models have been quite popular, and easier
to treat, first-principles calculations \cite{Andersen} indicate that such
models are probably oversimplified. Here an approach is proposed to the
cuprates, taking into account the existence of both ``large-$U$'' and
``small-$U$'' orbitals in the vicinity of the Fermi level ($E_{_{\rm
F}}$). 

\section*{AUXILIARY PARTICLES}

The large-$U$ orbitals are treated using the ``slave-fermion'' method
\cite{Barnes}. An electron of these orbitals at site $i$ and of spin
$\sigma$ is then created by $d_{i\sigma}^{\dagger} = e_i^{\dagger}
s_{i,-\sigma}^{\dagg}$, if it is in the ``upper-Hubbard-band'', and by
$d_{i\sigma}^{\prime\dagger} = \sigma s_{i\sigma}^{\dagger}
h_i^{\dagg}$, if it is in a Zhang-Rice-type ``lower-Hubbard-band''. Here
$e_i^{\dagg}$ and $h_i^{\dagg}$ are ``excession'' and ``holon'' fermion
operators, and $s_{i\sigma}^{\dagg}$ are ``spinon'' boson operators.
These auxiliary particle operators should satisfy in each site the
constraint $ e_i^{\dagger} e_i^{\dagg} + h_i^{\dagger} h_i^{\dagg} +
\sum_{\sigma} s_{i\sigma}^{\dagger} s_{i\sigma}^{\dagg} = 1$. 

The constraint can be imposed on the average by introducing a
chemical-potential-like Lagrange multiplier. But the Hilbert space
(referred to as the auxiliary space) then contains many non-physical
states. However, since the time evolution of Green's functions is
determined by the Hamiltonian which obeys the constraint rigorously,
expressing Physical observables in term of Green's functions results in
a correct treatment of the physical subspace. This can be violated by
applying inappropriate approximations to the Green's functions. 

Within the ``spin-charge separation'' approximation two-particle
spinon-holon Green's functions are decoupled into products of
one-(auxiliary)-particle Green's functions. Such an approximation has
been shown to be appropriate in one dimension. 

The Bogoliubov transformation $s_{\sigma}^{\dagg}({\bf k}) =
\cosh{(\xi_{\sigma{\bf k}})} \zeta_{\sigma}^{\dagg}({\bf k}) +
\sinh{(\xi_{\sigma{\bf k}})} \zeta_{-\sigma}^{\dagger}(-{\bf k})$ is
applied to diagonalize the spinon states. The diagonalized operators
$\zeta_{\sigma}^{\dagger}({\bf k})$ create spinons of ``bare'' energies
$\epsilon^{\zeta} ({\bf k})$. These energies have a V-shape zero minimum
at ${\bf k}={\bf k}_0$, where ${\bf k}_0$ is either $( {\pi \over 2{\rm
a}} , {\pi \over 2{\rm a}} )$ or $( {\pi \over 2{\rm a}} , -{\pi \over
2{\rm a}} )$. Bose condensation results in antiferromagnetism (AF), and
the spinon reciprocal lattice is extended over the basic reciprocal
lattice by adding the vector ${\bf Q}=2{\bf k}_0$. 

\section*{STRIPES AND CARRIERS}

It has been shown \cite{Zaanen,Emery} that a lightly doped AF plane
tends to phase-separate into ``charged'' and AF regions (gaining both
hopping and exchange energies). A preferred structure under long-range
Coulomb repulsion is of stripes of these phases, at least on the short
range. Such a scenario is supported by experiment
\cite{Bian,Tran,stripes}. A structure of narrow charged stripes forming
antiphase domain walls between wider AF stripes, has been confirmed for
at least some cuprates, and there exists growing evidence indicating
that such a structure probably exists, at least dynamically on the short
range, in all the superconducting cuprates. 

Spin-charge separation applies along the charged stripes (being one
dimensional). Holons (excessions) within these stripes are referred to
as ``stripons''. They carry charge, but not spin. Their fermion creation
operators are denoted by $p^{\dagger}_{\mu}({\bf k})$, and their bare
energies by $\epsilon^p_{\mu}({\bf k})$. Note that ${\bf k}$ here
corresponds to an approximate periodicity determined by the stripes
structure. 

It has been observed \cite{stripes} that the stripes in the cuprates are
quite ``frustrated'', and consist of disconnected segments. Since
itinerancy in one-dimension requires perfect order, it is assumed here
that an appropriate starting point is of localized stripon states. 

The effect of the small-$U$ orbitals is the existence of other carriers
(of both charge and spin) whose states are hybridized small-$U$ states
and those coupled holon-spinon and excession-spinon states which are
orthogonal to the stripon states. These carriers as referred to as
``Quasi-electrons'' (QE's). Their fermion creation operators are denoted
by $q_{\iota\sigma}^{\dagger}({\bf k})$. Their bare energies
$\epsilon^q_{\iota} ({\bf k})$ form quasi-continuous ranges of bands
crossing $E_{_{\rm F}}$ over ranges of the Brillouin zone (BZ). 

\section*{COUPLING VERTEX}

The auxiliary space fields are coupled to each other due to hopping and
hybridization terms of the original Hamiltonian. This coupling can be
expressed in terms of the following effective Hamiltonian term whose
parameters could be, in principle, derived self-consistently:
\begin{eqnarray}
{\cal H}^{\prime} &=& {1 \over \sqrt{N}} \sum_{\iota\mu\lambda\sigma}
\sum_{{\bf k}, {\bf k}^{\prime}} \Big\{\sigma
\epsilon^{qp}_{\iota\mu\lambda\sigma}({\bf k}^{\prime}, {\bf k})
q_{\iota\sigma}^{\dagger}({\bf k}) p_{\mu}^{\dagg}({\bf k}^{\prime})
\nonumber \\ &\ &\times\big[ \cosh{(\xi_{\lambda\sigma,({\bf k} - {\bf
k}^{\prime})})} \zeta_{\lambda\sigma}^{\dagg}({\bf k} - {\bf
k}^{\prime}) \nonumber \\ &\ &+ \sinh{(\xi_{\lambda\sigma,({\bf k} -
{\bf k}^{\prime})})} \zeta_{\lambda,-\sigma}^{\dagger}({\bf k}^{\prime}
- {\bf k}) \big] + h.c. \Big\}, 
\end{eqnarray} 

\begin{figure}[b!] % fig 1
\centerline{\epsfig{file=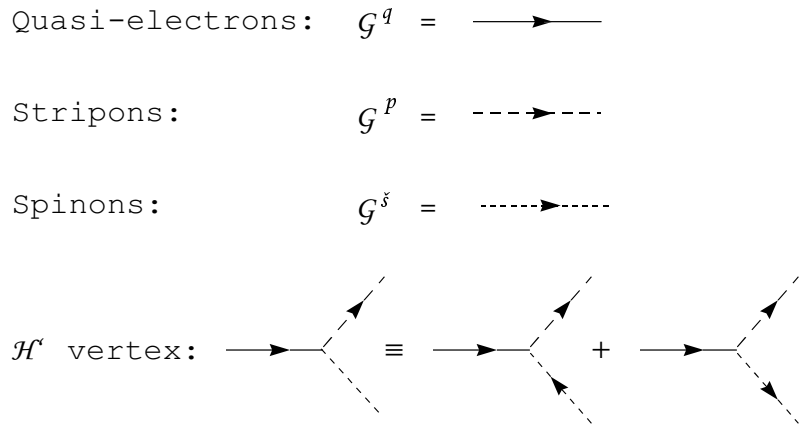,height=2.5in,width=4.25in}}
\vspace{10pt}
\caption{Diagrams for the auxiliary space propagators and the coupling 
vertex between them.} 
\label{fig1}
\end{figure}

Let us denote the QE, stripon, and spinon Green's functions by ${\cal
G}^q$, ${\cal G}^p$, and ${\cal G}^{\zeta}$, respectively. The
propagators corresponding to them are presented diagrammatically in
Fig.~1. ${\cal H}^{\prime}$ introduces a coupling  vertex between these
propagators, as shown in Fig.~1 too. As will be discussed below, the
stripon bandwidth turns out to be at least an order of magnitude smaller
than the QE and spinon bandwidths. Consequently one gets using a
generalized Migdal theorem that ``vertex corrections'' are negligible.

In Fig.~2 are presented diagrammatically the self-energy corrections
$\Sigma^q$, $\Sigma^p$, and $\Sigma^{\zeta}$, obtained for the QE's,
stripons, and spinons, respectively. 

\begin{figure}[b!] % fig 2
\centerline{\epsfig{file=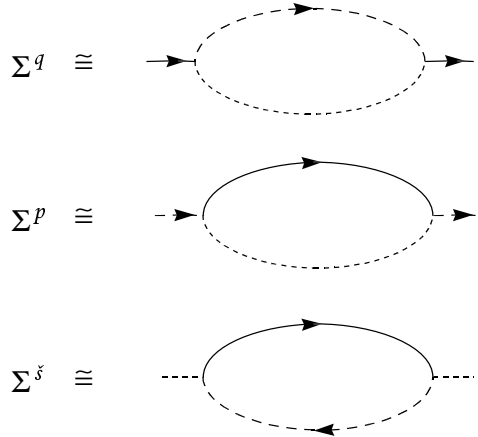,height=2.5in,width=2.75in}}
\vspace{10pt}
\caption{Diagrams for the self-energy corrections of the auxiliary space 
fields} 
\label{fig2}
\end{figure}

The auxiliary space spectral functions $A({\bf k}, \omega) \equiv \Im
{\cal G}({\bf k}, \omega-i0^+)/ \pi$, and scattering rates $\Gamma({\bf
k}, \omega) \equiv 2\Im \Sigma({\bf k}, \omega-i0^+)$ are denoted by
$A^q$, $A^p$, and $A^{\zeta}$, and by $\Gamma^q$, $\Gamma^p$, and
$\Gamma^{\zeta}$, for  the QE's, stripons, and spinons, respectively. 

\section*{QUASIPARTICLES}

For sufficiently doped cuprates the self-consistent self-energy
corrections determine quasiparticles of the following features: 

\subsection*{Spinons}

One gets spinon spectral functions behaving as: $A^{\zeta}({\bf k},
\omega)\propto\omega$ for small $\omega$. Consequently $A^{\zeta}({\bf
k}, \omega) b_{_T}(\omega)\propto T$ for $\omega\ll T$, where
$b_{_T}(\omega)$ is the Bose distribution function (at temperature $T$).
Thus there is no long-range AF order (associated with the divergence in
the number of spinons at ${\bf k}={\bf k}_0$).

\subsection*{Stripons}

The energies of the localized stripon states are renormalized to a very
narrow range around zero, thus getting polaron-like states. Some hopping
via QE-spinon states results is the onset of coherent itineracy at low
temperatures, with a bandwidth of $\sim$$0.02\;$eV. The stripon
scattering rates can be expressed as:
\begin{equation}
\Gamma^p({\bf k}, \omega) \propto A \omega^2 + B \omega T + CT^2. 
\end{equation}

\subsection*{Quasi-electrons}

The QE scattering rates, can be approximately expressed as: 
\begin{equation}
\Gamma^q({\bf k}, \omega) \propto \omega[b_{_T}(\omega) + \half], 
\end{equation}
becoming $\Gamma^q({\bf k}, \omega)\propto T$ in the limit $T\gg
|\omega|$, and $\Gamma^q({\bf k}, \omega)\propto\half |\omega|$ in the
limit $T\ll |\omega|$, in agreement with ``marginal Fermi liquid''
phenomenology \cite{Varma}. 

\subsection*{Phonon-dressed spinons (svivons)}

It was found \cite{Bian} that the charged stripes are characterized by
LTT structure, while the AF stripes are characterized by LTO structure.
The result would be that in any physical process induced by the ${\cal
H}^{\prime}$ vertex [see Eq. (1) and Fig. 1], the transformation of a
stripon into a QE, or vice versa, through the emission/absorption of a
spinon, is followed also by the emission/absorption of phonons. Thus the
stripons have also lattice features of polarons, and the spinons are
``dressed'' by phonons in processes induced by the ${\cal H}^{\prime}$
vertex. We refer to such a phonon-dressed spinon as a ``svivon'', and
its propagator can be expressed as a spinon propagator multiplied by a
power series of phonon propagators, as shown diagrammatically in Fig. 3.
The svivons carry spin, but not charge, however they also ``carry''
lattice distortion. 

\begin{figure}[b!] % fig 3
\centerline{\epsfig{file=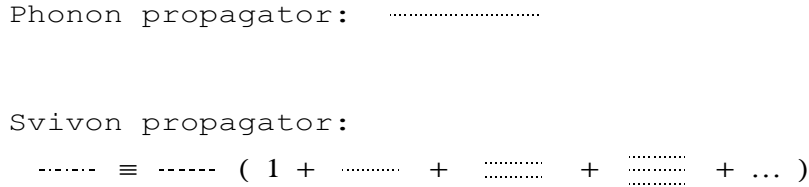,height=1.25in,width=4.25in}}
\vspace{10pt}
\caption{Diagrams for a phonon propagator and a svivon propagator
(expressed in terms of spinon and phonon propagators).} 
\label{fig3}
\end{figure}

\section*{SOME ANOMALOUS PHYSICAL PROPERTIES}

\subsection*{Optical conductivity}

The optical conductivity of the doped cuprates can be expressed
\cite{Tanner} as a combination of a Drude term and mid-IR peaks. Within
the present approach the Drude term results from transitions between low
energy QE states, while excitations of stripon states result in the
mid-IR peaks. Such excitations can either leave a stripon in the same
stripe segment, exciting spinon and phonon states, or transform it
through ${\cal H}^{\prime}$ into a QE and a svivon. 

\subsection*{Spectroscopic anomalies}

Experiments like photoemission give information about the electronic
spectral function, which is expressed as a combination of QE and
stripon-svivon contributions. Thus it has a ``coherent'' part, due to
the contributions of few QE bands, and an ``incoherent'' part of a
comparable weight, due to the contributions of other quasi-continuous QE
bands, and stripon-svivon states. 

The frequently observed $\sim$$|E-E_{_{\rm F}}|$ bandwidth is consistent
with Eq. (3). The spectroscopic "signature" of stripons is smeared over
few tenths of an eV around $E_{_{\rm F}}$ due to the accompanying svivon
excitations. The observed ``Shadow bands'' and ``extended'' van Hove
singularities result from the effect of the striped superstructure on
the QE bands \cite{Salk}. 

\subsection*{Transport properties}

\begin{figure}[b!] % fig 4
\centerline{\epsfig{file=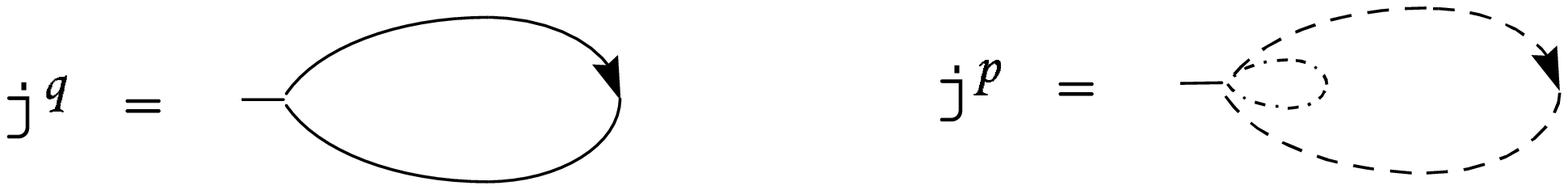,height=1.00in,width=4.00in}}
\vspace{10pt}
\caption{Diagrams for quasi-particles contributions to the electric
current.} 
\label{fig4}
\end{figure}

Within the present approach the electric current is expressed 
as a sum: ${\bf j} = {\bf j}^q + {\bf j}^p$, where the QE and stripon
contributions ${\bf j}^q$ and ${\bf j}^p$ are presented diagrammatically 
in Fig. 4. Since stripons transport occurs through
transitions to intermediate QE-spinon states, one gets ${\bf j}^p \cong
\alpha {\bf j}^q$, where $\alpha$ is approximately $T$-independent. In
order for this condition to be satisfied gradients $\bfnabla\mu^q$ and
$\bfnabla\mu^p$ of the QE and stripon chemical potentials must be formed
in the presence of an electric field or a temperature gradient, where
$N^q\bfnabla\mu^q + N^p\bfnabla\mu^p = 0$ ($N^q$ and $N^p$ are the
contributions of QE's and stripons to the electrons density of states at
$E_{_{\rm F}}$). 

Expressions for the dc conductivity and Hall constant are derived using
the Kubo formalism. Within the present approach they are expressed in
term if diagonal and non-diagonal conductivity QE terms
$\sigma_{xx}^{qq}$ and $\sigma_{xy}^{qqq}$, stripon terms
$\sigma_{xx}^{pp}$ and $\sigma_{xy}^{ppp}$, and mixed terms
$\sigma_{xy}^{qqpp}$. The diagrams for these terms are shown in Fig. 5. 

\begin{figure}[b!] % fig 5
\centerline{\epsfig{file=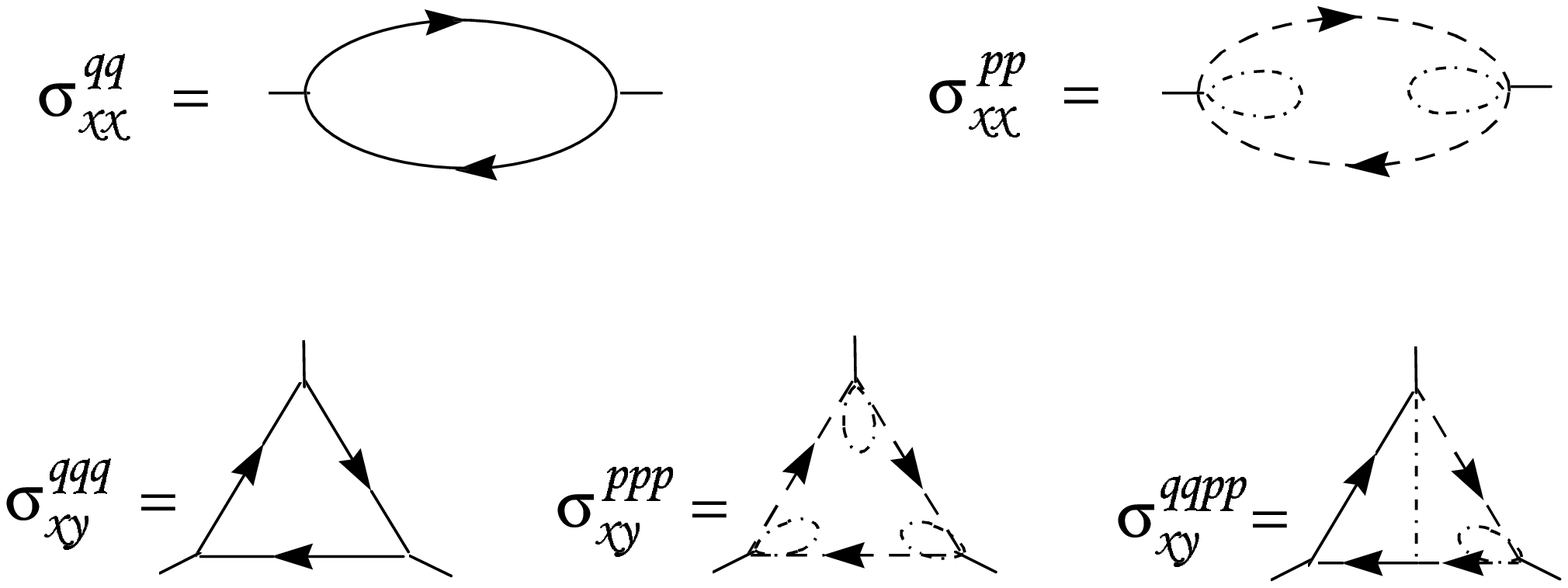,height=1.75in,width=4.25in}}
\vspace{10pt}
\caption{Diagrams for the QE, stripon, and mixed conductivity terms.} 
\label{fig5}
\end{figure}

It has been shown elsewhere \cite{Ashk} that the electrical resistivity 
can then be expressed as:
\begin{equation}
\rho_x = {1 \over (N^q+N^p) (1+\alpha)} \bigg( {N^q \over
\sigma_{xx}^{qq}} + {\alpha N^p \over \sigma_{xx}^{pp}}\bigg), 
\end{equation}
and the Hall constant as:
\begin{equation}
R_{_{\rm H}} = {\rho_x \over \cot{\theta_{_{\rm H}}}}, \ \ \ \
\cot{\theta_{_{\rm H}}} =  (1+\alpha) \bigg[{\sigma_{xy}^{qqq} +
\sigma_{xy}^{qqpp} \over \sigma_{xx}^{qq}} + {\alpha(\sigma_{xy}^{ppp} +
\sigma_{xy}^{qqpp}) \over \sigma_{xx}^{pp}} \bigg]^{-1}. 
\end{equation}

The temperature dependencies of these transport quantities are
determined by those of the scattering rates $\Gamma^q$ and $\Gamma^p$,
given in Eqs. (2), (3), to which temperature-independent impurity
scattering terms are added. Consequently one can express them in terms
of parameters $A$, $B$, $C$, $D$, $N$, and $Z$, as follows: 
\begin{eqnarray}
\sigma_{xx}^{qq} &\propto& {1 \over D+CT}, \ \ \ \ \ \ \ \ \ \
\sigma_{xx}^{pp} \propto {1 \over A+BT^2}, \nonumber \\
\sigma_{xy}^{qqq} \propto {1 \over (D+CT)^2}, \ \ &\ &\sigma_{xy}^{ppp}
\propto {1 \over (A+BT^2)^2}, \ \ \ \sigma_{xy}^{qqpp} \propto {1 \over
(D+CT)(A+BT^2)}. 
\end{eqnarray}
Resulting in the following expressions for $\rho_x$ and
$\cot{\theta_{_{\rm H}}}$: 
\begin{equation}
\rho_x = {D+CT+A+BT^2 \over N}, \ \ \ \ \cot{\theta_{_{\rm H}}} = \Big(
{Z \over D+CT} + {1 \over A+BT^2}\Big)^{-1}. 
\end{equation}
These expressions reproduce the systematic behavior of the transport
quantities in different cuprates, as has been demonstrated elsewhere
\cite{Ashk}. Note that one can get at the same time linear temperature
dependence of $\rho_x$ and quadratic temperature dependence of
$\cot{\theta_{_{\rm H}}}$, and that the temperature dependence of
$\rho_x$ can change to quadratic, and that of $\cot{\theta_{_{\rm H}}}$
to linear, as has been observed \cite{Chien}. 

Comparing the present analysis to that of Anderson \cite{Anderson}, who
first suggested that $\rho$ and $\cot{\theta_{_{\rm H}}}$ are determined
by different scattering rates (attributing the $T^2$ term to spinons),
it has been observed in ac Hall effect results \cite{Drew} that the
energy scale corresponding to the $T^2$ term is of $\sim$$120\;$K, which
is in agreement with energies of stripons (suggested here) and not of
spinons. 

It has also been shown elsewhere \cite{Ashk} that the thermoelectric
power (TEP) $S$ can be expressed in terms of QE and stripon terms $S^q$
and $S^p$, as: $S=(N^qS^q + N^pS^p)/(N^q + N^p)$, where $S^q\propto T$,
while the stripon term saturates at $T \simeq 200\; $K to $S^p=(k_{_{\rm
B}}/{\rm e})\ln{[(1}-n^p)/n^p]$, where $n^p$ is the fractional
occupation of the stripon band. 

This result is consistent with the typical behavior of the TEP in the
cuprates, and has been \cite{Tanaka} parametrized as: $S=AT+ BT^{\alpha}
/ (T+\Theta)^{\alpha}$. It was found \cite{Fisher,Mats} that $S^p=0$
(namely the stripon band is half full) for slightly overdoped cuprates. 

The effect of the doping is \cite{Tran} both to change the density of
the charged stripes within a CuO$_2$ plane, and to change the density of
carriers (stripons) within a charged stripe. It is the second type of
doping effect that changes $n^p$. 

\section*{PAIRING MECHANISM}

The ${\cal H}^{\prime}$ vertex provides a pairing mechanism which is
suggested here to drive high-$T_c$ superconductivity as well as the
normal-state pseudogap in the cuprates. This mechanism involves
transitions between pair states of QE's and stripons through the
exchange of svivons, as demonstrated diagrammatically in Fig. 6. It is
conceptually similar to the interband pair transition mechanism proposed
by Kondo \cite{Kondo}. The symmetry of the superconducting gap is
affected by {\bf k}-space symmetry which maximizes pairing. 

\begin{figure}[b!] % fig 6
\centerline{\epsfig{file=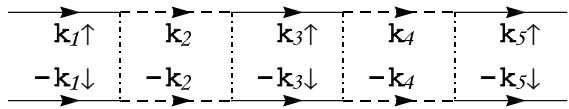,height=1.4in,width=3.25in}}
\vspace{10pt}
\caption{Diagram for transitions between pair states of QE's and
stripons, leading to pairing.} 
\label{fig6}
\end{figure}

A condition for superconductivity is that the narrow stripon band
maintains coherence between different stripe segments. The diagram
described in Fig. 6 can, however, drive pairing even when the stripons
are incoherent. If this occurs, the carriers do not carry supercurrent,
but a gap for pair breaking is still expected to exist. Such condensate
is interpreted here as the pseudogap phase found in underdoped cuprates.

Thus a normal-state psuedogap is expected to have a similar size and
symmetry to that of the superconducting gap, as has been observed
\cite{Norman}. Also the opening of the pseudogap should account for most
of the pair-condensation energy, as has been observed. 

If the BCS-like pairing temperature (below which the gap opens) is
denoted by $T_{\rm pair}$ and the stripon coherence temperature is
denoted by $T_{\rm coh}$, one expects superconducting transition at: 
\begin{equation}
T_c = \min{(T_{\rm pair}, T_{\rm coh})}.
\end{equation}
Thus  $T_c = T_{\rm coh} < T_{\rm pair}$ in underdoped cuprates, and
$T_c = T_{\rm pair} < T_{\rm coh}$ in overdoped cuprates, in agreement 
with the observed behavior of the gap \cite{mihail}.

Stripon coherence is energetically favorable at temperatures where there 
is a clear distinction between occupied and unoccupied stripon band
states. Thus, an estimate for $T_{\rm coh}$ for an almost empty (full)
stripon band is given by the distance ${\cal E}_{_{\rm F}}$ of the Fermi
level from the bottom (top) of the band at $T=0$. Using a
two-dimensional parabolic approximation one can express: 
\begin{equation}
k_{_{\rm B}} T_{\rm coh} \simeq {\cal E}_{_{\rm F}} = 2\pi\hbar^2 (n^* /
m^*), 
\end{equation}
where $m^*$ in the stripons effective mass and $n^*$ is their density
per unit area of a CuO$_2$ plane (note that the stripons are spinless). 

This result agrees with the ``Uemura plots'' \cite{Uemura} if the
$n^*/m^*$ ratio for stripons is approximately proportional to that for
the supercurrent carriers, appearing in the expression for the London
penetration depth. The ``boomerang-type'' behavior of the Uemura plots
in overdoped cuprates \cite{Nieder} is consistent with as a transition
from a band-top $T_c = T_{\rm coh}$ to a band-bottom $T_c = T_{\rm
pair}$ behavior, as discussed above. 

\section*{CONCLUSIONS}

The existence of high-$T_c$ superconductivity in the cuprates has been a
challenge for both experimentalists and theorists over the last 13
years. These complex materials have been found to be anomalous in almost
any physical property, and the traditional methods developed for simple
materials may be inadequate dealing with them. 

Here these materials are approached going beyond the ``standard''
models, and considering the effect of both large-$U$ and small-$U$
orbitals. A locally inhomogeneous striped structure is obtained, as well
as a non-standard existence of three types of carriers: polaron-like
stripons carrying charge, quasielectrons carrying charge and spin, and
svivons carrying spin and lattice distortion. 

Anomalous normal-state properties of the cuprates, and specifically
transport properties, are clarified, and a pairing mechanism based on
transitions between pair states of stripons and quasielectrons through
the exchange of svivons is derived, leading to high-$T_c$
superconductivity and to the normal-state pseudogap.

\end{document}